\documentclass[12pt,english,preprint]{aastex}
\usepackage{pslatex}
\usepackage[T1]{fontenc}
\usepackage[latin1]{inputenc}
\makeatletter

\shorttitle{SNRs in Molecular Clouds}
\shortauthors{Malkov {\it et al.~}}

\newcommand\jetp{{Sov. Phys. JETP\ }}	

\newcommand\rpp{{\it Rep.~Progr.~Phys.\ }}

\newcommand\etal{{\it et al.~}}
\newcommand\eg{{\it e.g.,~}}

\usepackage{babel}
\makeatother
\begin{document}

\title{On the gamma-ray spectra radiated by protons accelerated in SNR shocks
near molecular clouds: The case of SNR RX J1713.7-3946}

\author{M.A. Malkov\altaffilmark{1}, P.H. Diamond\altaffilmark{1} and
R.Z. Sagdeev\altaffilmark{2}}

\altaffiltext{1}{University of California at San Diego, La Jolla, California 92093-0319,
USA; E-mail: mmalkov@ucsd.edu}
\altaffiltext{2}{University of Maryland, College Park, Maryland 20742-3280, USA}

\begin{abstract}
Cosmic rays (CRs) are thought to be accelerated in SNRs. The most
favorable situation for proving that the main, hadronic CR component
is accelerated there is when CRs interact with dense gases, such as
molecular clouds (MC) which surround the SN shock. Here, a new mechanism
of spectrum formation in partially ionized gases near SNRs is proposed.
Using an analytic model of nonlinear diffusive shock acceleration,
we calculate the spectra of protons and estimate the resulting $\gamma$-ray
emission occurring when the SNR shock approaches a MC. We show that
the spectrum develops a break in the TeV range and that its GeV component
is suppressed. These modifications to the standard theory occur because
of the proximity of the partially ionized MC-gas and because of the
physics of particle and Alfven wave propagation inside the gas. Possible
applications of the new spectra to the recent CANGAROO and HESS observations
of the SNR RX J1713.7-3946 are discussed. 
\end{abstract}


\section{Introduction}

Cosmic rays (CRs) travel to us through the chaotic magnetic field
of the Galaxy. Therefore, the only way to connect them to their accelerator
is through the on-site radiation. Accelerated electrons have already
been detected in supernova remnant (SNR) shocks \citep{koyama95}.
However, electrons comprise only 1-2\% of the CR intensity above 4-5
GeV. Therefore, distinguishing between leptonic and hadronic origin
of the observed TeV emission from SNRs is a key to the proof of the
supernova origin of CRs. Simultaneous monitoring of X- and $\gamma-$
energy bands is an indispensable tool for that purpose. The same TeV
\emph{electrons} radiate in X- and $\gamma$-rays via synchrotron
and inverse Compton (IC) mechanisms, respectively \citep[\eg][]{sturn97,reyn}.
Accelerated \emph{protons}, can be detected \emph{only} near their
own (most interestingly TeV) energy band through their interaction
with an ambient gas. Therefore, the {}``smoking gun'' for \emph{proton
acceleration} should be the $\gamma$-ray emission, without an X-ray
emission that can be identified as the electron synchrotron radiation.
Molecular clouds (MC) adjacent to the SNR will dramatically enhance
proton visibility \citep[\eg][]{adv94,dav}.

However, to conclusively detect acceleration of super-TeV nuclei turned
out to be a very difficult task. Despite extensive search campaigns
\citep[\eg][]{buck98,vlk00}, only three SNRs have shown detectable
TeV emission so far \citep{tanim98,muaraishi,aharcasa,enom}. Nevertheless,
a signature of protons accelerated to super-TeV energies was reportedly
discovered in the SNR RX J1713.7-3946 by the CANGAROO team \citep{enom}.
Recently, this remnant has been confirmed as a TeV source, with significantly
reduced systematic and statistical errors in the range 1-10 TeV, by
the HESS collaboration \citep{aharnat}. This remnant is of a shell
type, typical for the major acceleration models, and rather abundant
in the Galaxy. Therefore, after some verification of the data and
analysis, it could serve as a direct evidence of the CR - SNR connection.

A number of groups \citep[\eg][]{buttnat,pohl02} reanalyzed the data
of \citet{enom} and claimed that the nucleonic interpretation is
{}``highly unlikely'' because of an alleged inconsistency with the
commonly assumed first order Fermi acceleration theory. We argue below
that if \citep[as it is assumed for this SNR  by][]{butt01,enom},
some of the accelerated particles begin to interact with a partially
ionized dense gas such as a MC in the \emph{northwestern} rim of the
remnant, the standard acceleration model is not applicable. In particular,
the nonlinearity of the acceleration process as well as Alfven wave
evanescence in the MC are essential. Therefore, the analyses by both
sides of the controversy are oversimplified and we believe that a
conclusive case for or against nucleonic origin of the emission has
not been made. Moreover, a recent analysis by \citet{pan} indicates
that the standard acceleration model applied to TeV electrons does
not adequately fit the TeV data, either, unless the X-ray emitting
filaments have an unreasonably low filling factor of $10^{-3}$. \citet{lazendic}
have increased it to $10^{-2}$ by allowing a smoother spectrum cut-off.

These analyses indicate that modifications to the accelerated particle
spectra are required. Namely, a low-energy cut-off above $100GeV$
and/or a \emph{spectral break} would support the nucleonic scenario
\citep{pohl02} while models with an exponential cut-off rather than
a break provide only a bad fit. Such modifications are suggested below.

The controversy about the TeV observations of the SNR RX J1713.7-3946
is fundamental and will clearly impair understanding of future observations
of SNRs nearby MCs. Therefore, in this Letter we revisit shock acceleration
theory and include the following physical phenomena in the model:
(i) the nonlinearity of the acceleration process, i.e., the modification
of the flow by accelerated particles (ii) a position dependent \emph{low-energy
cut-off} of accelerated particles ahead of the shock, which is \emph{well
known} to be present in analytic solutions for the shock acceleration
problem. Note that this cut-off is usually ignored in shock acceleration
models since only the downstream solution is considered. When the
dense gas is present upstream of the shock this cut-off should be
included in the calculation of $\gamma$-radiation. (iii) impairment
of CR confinement in a dense gas (MC) due to non-propagation of some
Alfven waves. The latter phenomenon produces a break (curvature) in
the particle and radiation spectra in the TeV energy range and, in
combination with (i), results in a spectral slope significantly different
from the usual predictions of linear theory supplemented with the
high energy cut-off, which was used in the analyses of \citet{enom}
and \citet{pohl02}, for example.

\section{Low-energy cut-off and the absence of thermal $x$-ray emission}

The diffusive shock acceleration (DSA) mechanism operates by scattering
charged particles off magnetic irregularities (Alfven waves) across
a shock, such as a SN blast wave \citep[\eg][]{be87}. In the linear
acceleration regime, the momentum distribution is a power-law, $p^{2}f_{0}(p)=Cp^{-s}$
where $C$ is a normalization constant and the index $s=\left(r+2\right)/(r-1)$
depends only on the shock compression ratio $r$. As soon as the flow
is disturbed by the pressure of accelerated particles, the spectrum
becomes more complex. It flattens toward higher energies and may become
steeper at lower energies. The spectrum $f_{0}\left(p\right)$ is
the \emph{downstream} spectrum, which is coordinate independent in
both the linear and nonlinear regimes. Here the upstream spectrum
is more important since the interaction with the adjacent MCs starts
\emph{upstream} where a low energy \emph{cut-off} occurs. Indeed,
the solution upstream, valid for both the linear and nonlinear regimes,
reads \citep[\eg][]{mdru01}

\begin{equation}
f(p,x)=f_{0}(p)\exp\left[-\frac{\alpha}{\kappa}\int_{0}^{x}udx\right].\label{spec}\end{equation}
 Here $f_{0}(p)$ is the spectrum downstream, $\alpha(p)\sim1$. The
$x$ coordinate points upstream ($x>0$) from the shock ($x=0$) and
$u(x)>0$ is the speed of the plasma flowing into the shock ($u(x)\sim V_{shock}$).
Now, $\kappa(p)$ grows with $p$, most likely linearly, $\kappa\approx\kappa'p$.
According to eq.(\ref{spec}), there exists a CR precursor, which
is of the length $l_{CR}=\kappa\left(p_{max}\right)/V_{shock}$. Furthermore,
high energy particles diffuse ahead of low energy particles so the
spectrum (\ref{spec}) has an exponential \emph{low-energy cut-off}
at

\begin{equation}
p_{min}(x)\approx\frac{1}{\kappa'}\int_{0}^{x}udx\sim V_{shock}x/\kappa'.\label{eq:pmin}\end{equation}
 For the Bohm diffusion we have

\begin{equation}
p_{min}/mc\approx\left(V_{shock}/c\right)B_{\mu}\left(x/10^{12}cm\right)\label{eq:pminB}\end{equation}
 where $B_{\mu}$ is the magnetic field in $\mu G$. This low-energy
decay can explain why the same MC can be visible in the TeV energy
range and invisible in the GeV range. Let the near-most edge of an
adjacent MC be upstream of the shock at a distance $x=x_{MC}$, such
that $1GeV/c<p_{min}(x_{MC})<p_{max}$. This low-energy cut-off $p_{min}$
is shown in Fig.\ref{fig:sp}%
\footnote{We have chosen the following values for the parameters in eq.(\ref{eq:pminB}):
$V_{shock}=1000km/sec$, $B_{\mu}=3$, and the distance from the sub-shock
to the MC, $x=x_{MC}\approx7\cdot10^{15}cm$. Note that for the visibility
of the TeV and higher energy protons, it is sufficient that $x_{MC}<10^{17}cm$.%
} that will be discussed below. The last condition means that the leading
edge of the CR precursor (filled with TeV protons) has already penetrated
into the MC, but the sub-shock itself (with the GeV particles ahead
of it) still has not. Obviously, the CANGAROO and HESS could detect
the proton TeV emission while the EGRET could not, because of the
low density inside the wind bubble the sub-shock is located in. The
strong X-ray emission expected when the sub-shock crashes into the
dense MC will not yet be visible, either.

\section{High-energy spectral break}

In the presence of weakly ionized dense gas the particle and emission
spectra undergo significant modifications in the TeV energy range,
where a spectral break can form. According to Eq. (\ref{fig:sp})
the accelerated particles occupy an extended precursor of the size
\begin{equation}
l_{CR}\sim\kappa(p_{max})/V_{shock}>r_{g}(p_{max})c/V_{shock}\label{eq:lcr}\end{equation}
 ahead of the shock. Here $r_{g}$ is the particle gyroradius. One
can estimate $l_{CR}$ as \\
 $l_{CR}\sim\left(p_{max}/mc\right)\left(c/V_{shock}\right)B_{\mu}^{-1}10^{12}cm.$
Thus, the shock precursor may be as long as $10^{18}cm$ for $p_{max}\sim10TeV$.
Therefore, the accelerated particles start to interact with a MC before
it becomes significantly ionized by the shock wave or the ionizing
precursor \citep{Draine93}. Hence, they propagate in the MC under
conditions of strong ion-neutral collisional damping of the self-generated
Alfven waves that are needed to confine accelerated particles. More
importantly, there is a gap in wavenumber space at $k_{1}<k<k_{2}$
where the waves do not propagate \citep{kulsr69,zweib82}. Here, it
is critical to realize that waves \emph{literally do not exist}, as
opposed to simply being damped by collisions. The above wave evanescence
range is bounded by $k_{1}=\nu_{in}/2V_{A}$ and $k_{2}=2\sqrt{\rho_{i}/\rho_{0}}\nu_{in}/V_{A}$,
where $\rho_{i}/\rho_{0}\ll1$ is the ratio of the ion to neutral
mass density, $\nu_{in}$ is the ion-neutral collision frequency and
$V_{A}=B/\sqrt{4\pi\rho_{i}}$ is the Alfven speed. The resonance
condition for the wave generation and particle scattering off them
is $kp_{\parallel}/m=\pm\omega_{c}$, where $\omega_{c}=eB/mc$ and
$p_{\parallel}$ is the parallel (to the magnetic field) component
of particle momentum. Therefore, particles having $p_{1}<\left|p_{\parallel}\right|<p_{2}$,
where

\begin{equation}
p_{1}=2V_{A}m\omega_{c}/\nu_{in},\;\; p_{2}=\frac{1}{4}\sqrt{\frac{\rho_{0}}{\rho_{i}}}p_{1}>p_{1},\label{eq:p12}\end{equation}
 are not scattered by any waves, so that the confinement of all particles
with $\left|p_{\parallel}\right|>p_{1}$ is dramatically degraded.
We assume that the gap in $p_{1}<p_{\parallel}<p_{2}$ is sufficiently
broad, so that resonance broadening \citep[\eg][]{acht81} does not
bridge it. We also assume that there is no significant background
turbulence at the scales $\lambda>2\pi/k_{1}$, that could reduce
particle diffusivity to a limit which is not significantly higher
than the Bohm value. For the parameters used in our calculations below,
$k_{1}\sim10^{-14}cm^{-2}$. Then, particles with $\left|p_{\parallel}\right|>p_{1}$
will escape from the MC upstream of the shock along the magnetic field
at a speed comparable to $c$ or, at least, their confinement time
to the shock will be much shorter than those with $\left|p_{\parallel}\right|<p_{1}$.
Only the latter particles will generate $\pi^{0}$ mesons and $\gamma$
emission efficiently in the MC. Their pitch angle distribution for
$p>p_{1}$ is, however, limited to the interval $\left|\mu\right|<p_{1}/p$.
Therefore, the contribution to the $\gamma$ emission of particles
with momentum $p>p_{1}$ is reduced by the phase volume filling factor
$\mu_{crit}=p_{1}/p$. The resulting emission spectrum is thus one
power steeper as compared to the standard calculations \citep[\eg][]{berptus,dav,naito}
based on the isotropic particle distribution. The latter yields the
$\gamma$ spectrum which reproduces that of the particles up to about
$0.1cp_{max}$ where it declines. Hence, the energy spectrum of $\gamma$
emission must be the same as that of the particles, $\propto\varepsilon_{\gamma}^{-s}$,
for the photon energy $\varepsilon_{\gamma}<\varepsilon_{br}\sim0.1cp_{1}$
(particle momenta $p<p_{1}$), and it must scale as $\propto\varepsilon_{\gamma}^{-s-1}$
for $\varepsilon_{\gamma}\ga\varepsilon_{br}$ up to about $0.1cp_{max}$
(particle momenta $p_{1}<p<p_{max}$). The break energy $\varepsilon_{br}$
can be estimated using Eq. (\ref{eq:p12}) for $p_{1}$ by substituting
$\nu_{in}=n_{0}\left\langle \sigma V\right\rangle $, where $\sigma$
is the cross-section of the ion-neutral collisions and $V$ is the
collision velocity averaged over a thermal distribution. Using an
approximation of \citet{kulsr71} for $\left\langle \sigma V\right\rangle $,
$p_{1}$ can be estimated as \begin{equation}
p_{1}/mc\simeq16B_{\mu}^{2}T_{4}^{-0.4}n_{0}^{-1}n_{i}^{-1/2}.\label{eq:p1}\end{equation}
 where $T_{4}$ is measured in the units of $10^{4}K$, $n_{0}$ and
$n_{i}$ (number densities corresponding to the mass densities $\rho_{0}$
and $\rho_{i}$) \--in $cm^{-3}$.

To illustrate these results we consider three different acceleration
regimes, corresponding to the three substantially different shock
compression ratios $r$, shown by the points 1-3 in the inset to Fig.\ref{fig:sp}.
(Here $r$ means a total shock compression, that is adiabatic compression
across the precursor times the subshock jump.) As it can be seen from
this plot, the shock compression is difficult to calculate accurately
due to its sharp dependence on the injection rate (usually not well
known) and other parameters, such as the maximum momentum. We discuss
this spectrum in more detail in the next section.

\section{Discussion and Conclusions}

In this paper we have considered the \emph{upstream} particle spectrum
since this is necessary for interpretation of cases where a shock
is expanding into a low-density pre-supernova wind bubble and is approaching
a denser material such as a swept-up shell or a MC. Perhaps the most
discussed SNR of this kind is the RX J1713.7-3946 claimed by the CANGAROO
team to be a long anticipated proton super-TeV accelerator \citep{enom}.
Significant part of the TeV emission detected by the CANGAROO and
HESS \citep{aharnat} instruments comes from the northwestern rim
of the remnant where the interaction with a molecular cloud is believed
to take place \citep[\eg][]{slane,hiraga}. Therefore these observations
are pertinent to the subject of this letter.

\begin{figure}
\plotone{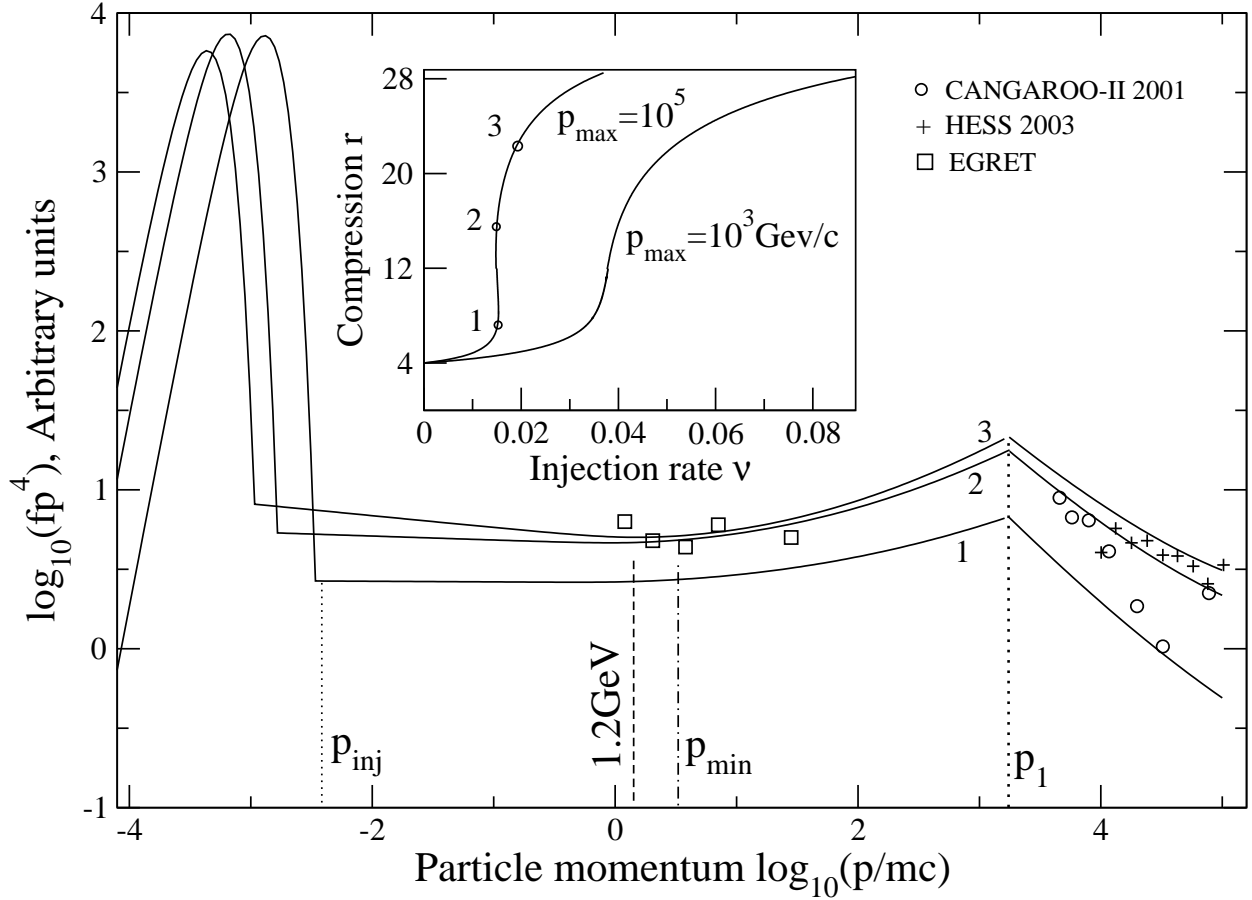}

\caption{Inset: response of the shock structure to the particle acceleration.
The flow compression ratio $r$ is shown as a function of particle
density $n_{CR}$ injected into the acceleration from the thermal
distribution calculated for two maximum momenta $p_{max}$ and the
shock Mach number M=80. The injection parameter $\nu\sim(cp_{inj}/mV_{shock}^{2})n_{CR}/n_{1}$,
where $n_{1}$ is the plasma density upstream of the shock. The details
of these calculations and those of the spectra shown in the main figure
can be found in \citep{mdru01}. Main figure: spectra of accelerated
particles behind the shock front, for $p_{max}/mc=10^{5}$. Three
solutions correspond to the points 1-3 on the compression-injection
diagram in the inset. Note that the test particle (linear) spectrum
would be a horizontal line. The break energy (momentum) is at $p/mc=1.8\cdot10^{3}$.
The CANGAROO-II \citep{enom} and HESS points are adopted from \citet{aharnat}
while EGRET \citep{egr99} points-- from \citet{pohl02}. Vertical
lines indicate: the injection momentum (separating thermal and nonthermal
particles), the 1.2 GeV proton energy threshold of $\pi^{0}$ production,
the possible low-energy cut-off $p_{min}(x)$ and the spectral break
$p_{1}$ (see text).\label{fig:sp}}
\end{figure}

To demonstrate that the observed TeV spectra are consistent with the
mechanisms suggested in this paper, we put the calculated upstream
\emph{proton} spectra on one plot with the CANGAROO/HESS and EGRET
\emph{photon} fluxes that would be radiated by these protons. Since
the emission generated by protons has typically an order of magnitude
higher energy, we up-shifted the \emph{observed} emission energy accordingly
in Fig. \ref{fig:sp}. As the spectrum between $GeV/c<p<p_{1}$ is
considerably flatter than in the case of the linear theory, its magnitude
at $p=p_{1}$ is an order of magnitude higher for the same injection
rate. This lowers the ambient gas density required to produce the
same TeV emission as computed by \citet{enom} using the linear spectrum.
In addition, this allows to accommodate the EGRET points, should they
belong to the same object \citep{pohl02}. However, because of the
controversy about the distance to the remnant (6 vs 1 kpc) \citep{slane,fukui}
and other uncertainties, it is impossible to present detailed normalized
fits of the photon flux in this short letter. 

The form of the spectrum is marked by increasing the spectral index
by one at the break momentum $p=p_{1}$, Fig. \ref{fig:sp}. This
is crucial to fitting the observed spectra without imposing a specific
form and value of the energy cut-off. The latter procedure would give
only a poor fit \citep[\eg][]{pohl02}. Note that there are two CANGAROO
points that are not in a good agreement with both the HESS and with
the theoretical spectrum. It should be noted, however, that any disagreement
between the three sets of spectral points is substantially enhanced
in this particular spectrum format (the particle phase density is
multiplied by $p^{2}$). Except for these two points the agreement
is remarkably good. Interestingly enough the agreement is excellent
for a subset of HESS points with the lowest and highest energy points
excluded.

Unlike the power-law index, it is more difficult to constrain the
parameters determining the position of the break on the spectrum given
by Eq. (\ref{eq:p1}) due to the poor information about the target
gas for $p-p$ reactions (MC). For the particular case shown in Fig.\ref{fig:sp},
we have chosen such combination of parameters in Eq. (\ref{eq:p1})
that $p_{1}$ amounts to $p_{1}\sim1.8TeV/c$. MCs in general are
known to be {}``clumpy'' with the inter-clump gas density of 5-25
$cm^{-3}$ and a less than 10\% filling factor \citep[see, \eg][and references therein]{chevMC,bykov00}
. The results shown in Fig. \ref{fig:sp} are obtained for the following
values of parameters in eq.(\ref{eq:p1}): $B_{\mu}\approx14$, $T_{4}\approx0.01$,
$n_{0}\approx23$, and $n_{i}\approx0.23$ which yields $p_{1}/mc\approx1.8\cdot10^{3}$.
By contrast, in the model of \citet{ellis} much lower target density
is assumed since its authors do not consider the possibility that
the CR precursor reach the dense gas while the shock itself is still
in the wind bubble. Therefore they concluded that the $p-p$ reactions
do not contribute significantly to the TeV emission because of the
lack of target protons. In contrast to the conclusion of \citet{ellis}
and to that of \citet{enom} the most recent HESS observations allow
their authors \citep{aharnat} to assume that both protons and electrons
contribute to the TeV emission significantly.

Note that a complementary mechanism of suppression of the low-energy
(GeV) emission has been suggested earlier by \citet{ahat99} and discussed
recently in the context of SNR RX J1713.7-3946 by \citet{uchi}. This
mechanism requires an impulsive release of accelerated particles at
some distance from a MC (target) and subsequent (energy dependent)
diffusive propagation of CRs to the target. By contrast, the mechanism
discussed in this paper is based on a quasi-stationary solution of
the acceleration problem (Eq. {[}\ref{spec}{]}) which implies that
the accelerated particles are bound to the propagating shock front
via self-generated Alfven waves. It is clear, however, that in both
cases the suppression of the low-energy emission is based on a slower
diffusion of low-energy particles. Note that \citet{lazendic} also
discussed qualitatively the same effect. It is also important to mention
that GeV emission may not be suppressed all over the remnant, and
was demonstrated by \citet{butt01} to be likely of proton origin.
Obviously, the spectral modifications considered in this letter are
equally applicable to electrons of similar rigidity.

It should be pointed out that our model in its current version does
not provide a complete fit to the data, particularly in the X- and
radio energy bands. The TeV-GeV fluxes are given in arbitrary units,
so only the \emph{form} of the spectrum (including possible GeV signal)
is provided here. A complete fit would require the convolution of
the proton spectrum with appropriate emissivities given the densities
and the distance. The thermal peak in Fig.\ref{fig:sp} is a Maxwellian,
approximately normalized relative to the high-energy part of the spectrum,
in accordance with the calculated downstream temperature without turbulent
heating. More accurate calculation of the thermal peak would require
adequate injection model and turbulent gas heating in the precursor
\citep[\eg][]{byk96,mdru01,kjg02}. Such calculations of the thermal X-ray
emission should be tested specifically against very low observed upper
limits. One \emph{potentially} observable prediction of our model
is the break in the pitch angle averaged particle spectrum at $p=p_{1}$
that would be produced by almost instant losses along the field lines
of all particles with $\left|p_{\parallel}\right|>p_{1}$ once they
enter a partially ionized gas with significant ion-neutral collisions.
This break may very well be in the currently {}``obscure'' energy
range between the EGRET and the ground based Cherenkov telescope energy
bands. The next generation of space gamma telescopes, such as GLAST,
will be able to explore this energy range.

The principal results of this letter, which are the spectrum softening
by one degree above the spectral break, and the possibility of a low
energy cut-off clearly show that these physical phenomena need to
be included in the models to conclusively differentiate between the
nucleonic and leptonic sources of the TeV emission from the remnants
nearby molecular clouds such as the RX J1713.7-3946. Previous claims
to the contrary have ignored necessary refinements in the DSA theory
beyond the level of test particle theory.

\acknowledgements{We thank Felix Aharonian and Heinz V\"{o}lk for helpful discussions
as well as an anonymous referee provided useful comments on the manuscript.
Support under NASA grant ATP03-0059-0034 and under U.S. DOE grant
No. FG03-88ER53275 is gratefully acknowledged. }

\newpage

\end{document}